\documentclass[twocolumn,floatfix,showpacs,preprintnumbers,amsmath,amssymb]{revtex4}

\usepackage{graphicx}% Include figure files
\usepackage{dcolumn}% Align table columns on decimal point
\usepackage{bm}% bold math

\begin{document}
\preprint{APS/123-Neutrino Interactions}
\title{Nuclear Effects in Neutrino Induced Coherent Pion Production at K2K and MiniBooNE Neutrino Energies}
\author{S. K. Singh, M. Sajjad Athar}
\author{Shakeb Ahmad}
\affiliation{Department of Physics,
 Aligarh Muslim University, Aligarh- 202002, India.}

\begin{abstract}
The coherent pion production induced by neutrinos in nuclei is studied using a delta hole model in local density approximation taking into account the renormalization of $\Delta$ properties in a nuclear medium. The pion absorption effects have been included in an eikonal approximation. These effects give a large reduction in the total cross section. The numerical results for the total cross section are found to be consistent with recent experimental results from K2K and MiniBooNE collaborations and other older experiments in the intermediate energy region.
\end{abstract}


\maketitle
The present experiments at K2K and MiniBooNE are designed to search for $\nu$-oscillations in $\nu_\mu$ disappearance and $\nu_e$ appearance channels. In these experiments, the $\nu_\mu$-spectrum is determined by the observed energy spectrum of muons which are predominantly produced in forward direction through the charged current quasielastic reactions induced by $\nu_\mu$. In this kinematic region the major background to these events come from the non quasielastic events in which pions are produced through coherent and incoherent processes induced by charged and neutral weak currents in $\nu$-nucleus interactions. The neutral current induced $\pi^0$ production is of particular importance as it constitutes a major background to the electron signal in $\nu_e$ appearance channel. The analysis of $\nu$-oscillation experiments, therefore, requires a better understanding of pion production processes in $\nu$-nucleus interaction and many experiments are being done to study $\nu$-induced coherent and incoherent production of pions from nuclei~\cite{nuint}. 

The coherent production of charged pions from $^{12}\text{C}$ has been recently studied by the K2K collaboration~\cite{hasegawa}. The coherent production of neutral pions has been studied by the K2K collaboration for $^{16}\text{O}$~\cite{nakayama} and by the MiniBooNE collaboration for $^{12}\text{C}$~\cite{raaf}. Many experimental groups have earlier studied the coherent production of pions induced by neutrinos from other nuclei at higher energies~\cite{faissner}-\cite{marage}. Theoretically various authors~\cite{adler}-\cite{paschos} have used Adler's PCAC theorem to predict the total cross sections for neutrino reactions, which overestimate the experimental cross sections at low energies. In these calculations the nuclear medium effects are included only through the final state interaction of the outgoing pions with the nucleus using a model of pion nucleus scattering. A theoretical framework for treating the nuclear medium effects in the neutrino production of coherent pions using some model of nuclear structure has been recently discussed by some groups~\cite{kim} but no definite predictions are made for the kinematics of the neutrino oscillation experiments of present interest.

We present in this letter a calculation of the neutrino induced production of coherent pions from nuclei at intermediate neutrino energies. The nuclear medium effects are taken into account in the weak production process as well as in the final state interaction of the outgoing pions with the nucleus. The calculation uses the local density approximation to the delta hole model which has been used earlier to study photo and electro production of pions from nuclei~\cite{carrasco}. The final state interaction of pions has been treated in eikonal approximation with the pion optical potential described in terms of the self energy of a pion in a nuclear medium calculated in this model~\cite{oset1}.

In the kinematical region between threshold and few hundred MeV excitation energy relevant for the production of pions in neutrino oscillation experiments at intermediate neutrino energies, the dominant contributions are due to the excitation of the $\Delta$ resonance. We treat the $\Delta$ as a Rarita Schwinger field $\Psi_\lambda$ and use the on shell form~\cite{line} of the $\Delta$ propagator given by 
{\small
\begin{equation}
\Delta_{\lambda\sigma}(\text P)=\frac{\not\text{P}+\text M_\Delta}{\text P^2-\text M^2_\Delta+i\Gamma \text M_\Delta} {\cal P}^{3/2}_{\lambda\sigma}(\text P)
\end{equation}}
where $\Gamma$ is the energy dependent width of $\Delta$ and 
%${\cal P}^{3/2}_{\lambda\sigma}(\text P)$ is the $\Psi_\lambda$ field. 
${\cal P}^{3/2}_{\lambda\sigma}(\text P)$ is the spin-$\frac{3}{2}$ projection operator. 
We use ${\cal L}^{int}_{\Delta\text{N}\pi}=\frac{f_{\Delta \text N \pi}}{\text m_\pi}\bar\Psi_\lambda \partial^{\lambda}\phi \Psi + \text{h.c.}$ with $\frac{f^2_{\Delta \text N \pi}}{4\pi}$=0.36 to describe the on shell $\Delta\text N\pi$ interaction~\cite{line}-\cite{post} and the standard model of weak interactions to describe the $\Delta$-N isovector charge and neutral transition currents corresponding to W and Z exchanges. The matrix element for the charge current $\pi^+$ production from the proton, is calculated using the Feynman diagrams shown in Fig.~1, with the $\Delta$ propagator given in eq.~(1) and the $\text p\rightarrow\Delta^{++}$ weak transition current $J^\mu_\text W\text(s)$ in the s-channel, given by 
{\small
\begin{widetext}
\begin{eqnarray}
J^\mu_\text W\text(s)&=&{\sqrt 3}\bar{\text u}_\alpha(\text P)\left[\left(\frac{\text C^\text V_{1}(q^2)}{\text M}(g^{\alpha\mu}{\not q}-q^\alpha{\gamma^\mu})+\frac{\text C^\text V_{2}(q^2)}{\text M^2}(g^{\alpha\mu}q\cdot{\text P}-q^\alpha{\text P^{\mu}})
+\frac{\text C^\text V_3(q^2)}{\text M^2}(g^{\alpha\mu}q\cdot \text p-q^\alpha{\text p^\mu})\right)\gamma_5\right.\nonumber\\
&+&\left.\left(\frac{\text C^\text A_{1}(q^2)}{\text M}(g^{\alpha\mu}{\not q}-q^\alpha{\gamma^\mu})+\frac{\text C^\text A_{2}(q^2)}{\text M^2}(g^{\alpha\mu}q\cdot{\text P}-q^\alpha{\text P^{\mu}})
+\text C^\text A_{3}(q^2)g^{\alpha\mu}+\frac{\text C^\text A_4(q^2)}{\text M^2}q^\alpha q^\mu\right)\right]\text u(\text p)
\end{eqnarray}
\end{widetext}}
and a similar isospin rotated $\text p\rightarrow\Delta^0$ transition current $J^\mu_\text W\text(u)$ for the u-channel diagram. ${\text u_\alpha}(\text P)$ and u(p) are the Rarita Schwinger and Dirac spinors for $\Delta$ and nucleon.
\begin{figure}[ht]
\includegraphics{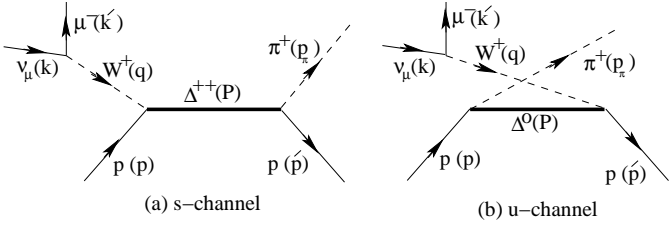}
\caption{Feynman diagrams}
\end{figure}
The form factors $\text C^\text V_j$(j=1-3) and $\text C^\text A_j$(j=1-4) are taken from the experimental analysis of neutrino production of pions from proton and deuteron targets~\cite{barish}-\cite{kitagaki}. The $q^2$ dependence of these form factors has been discussed in the literature by many authors~\cite{sch}-\cite{sakuda}. We use a dipole parametrization for the vector form factors and a modified dipole parameterization for the axial vector form factors given in Refs.~\cite{barish},~\cite{sch} with a vector dipole mass $\text M_\text V=\text{0.73~GeV}$ and an axial vector dipole mass $\text M_\text A=\text{1.05~GeV}$.

The transition current for the charged pion production from neutron targets is calculated in a similar manner using  the $\text n\rightarrow\Delta^+$ transition current for the s-channel and $\text n\rightarrow\Delta^-$ for the u-channel diagrams at the weak vertex. For the neutral current induced reaction, we obtain the transition current in a similar manner using the $\text p\rightarrow\Delta^+$ transition current for the production of $\pi^0$ from protons and the $\text n\rightarrow\Delta^0$ transition current for the production of $\pi^0$ from neutrons in s and u channels. With these considerations, the nuclear transition current ${J}^\mu_\text {W(Z)}$ for the coherent charged (neutral) pion production from a nucleus $^\text{A}X$ is obtained by summing over all the occupied nucleons in the amplitude and is written as 
{\small
\begin{equation}
{J}^\mu_\text {W,Z}=\sum_\text{i=s,u}{\cal T}^{\mu}_{\text W,Z}(i)\frac{\text M^2}{\text P^2_i-\text M^2_{\Delta}+i\Gamma \text M_\Delta}\int\rho^i_\text{W,Z}(r)e^{i({\vec q}-{\vec p_\pi})\cdot{\vec r}}\text d{\vec r}
\end{equation}}
$\text P_i$ is the momentum of the $\Delta$ resonance in i(=s,u) channel. ${\cal T}^{\mu}_\text {W,Z}(i)$ is the nonpole part of the kinematic factors involving transition form factors $\text C^\text {V,A}_j(q^2)$, $\rho^i_\text{W,Z}(r)$ is the linear combination of proton and neutron densities incorporating the isospin factors for charged and neutral pion production from proton and neutron targets corresponding to W and Z exchange diagrams in i(=s,u) channel. 
\begin{figure}[ht]
\includegraphics{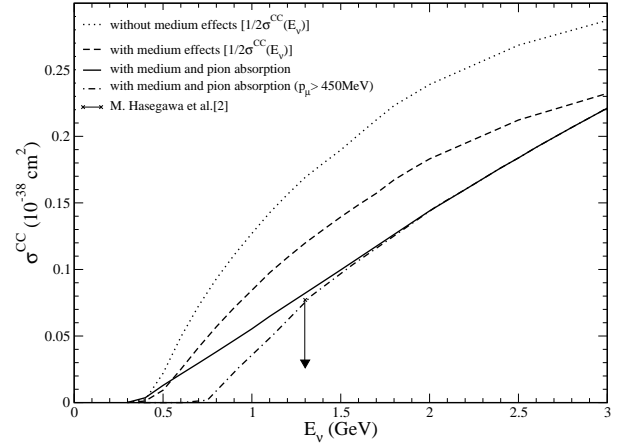}% Here is how to import EPS art
\caption{$\sigma(\text{E}_{\nu})$ vs $\text{E}_{\nu}$ for coherent $\pi^+$ production in $^{12}\text{C}$ (see text for details).}
\end{figure}
\begin{figure}[ht]
\includegraphics{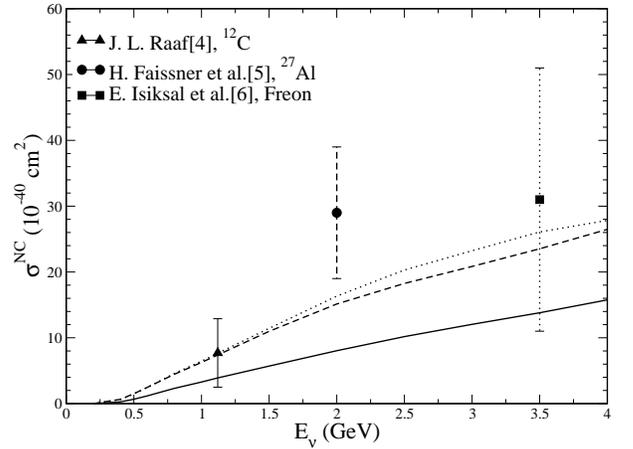}% Here is how to import EPS art
\caption{$\sigma$$(\text{E}_\nu)$ vs $\text{E}_{\nu}$ for coherent $\pi^0$ production in $^{12}\text{C}$(solid), $^{27}\text{Al}$(dashed) and Freon(dotted) with nuclear medium and pion absorption effects, along with the experimental results~\cite{raaf}-\cite{isiksal}}
\end{figure}
\begin{figure}
\includegraphics{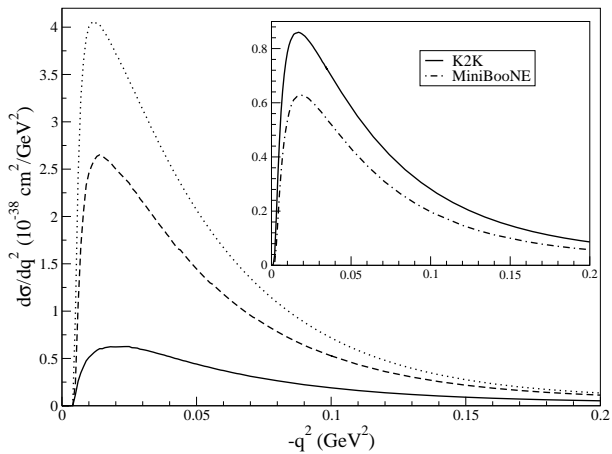}% Here is how to import EPS art
\caption{ $\frac{d\sigma}{dq^2}~ vs~ -q^2$ at $E_\nu$=1~GeV, for coherent $\pi^+$ production in $^{12}\text{C}$ nucleus without(dotted), with(dashed) nuclear medium effects and with nuclear medium and pion absorption effects(solid). In the inset the final results for $<\frac{d\sigma}{dq^2}$ vs $-q^2$ averaged over the K2K and MiniBooNE spectrum are shown.}
\end{figure}
\begin{figure}
\includegraphics{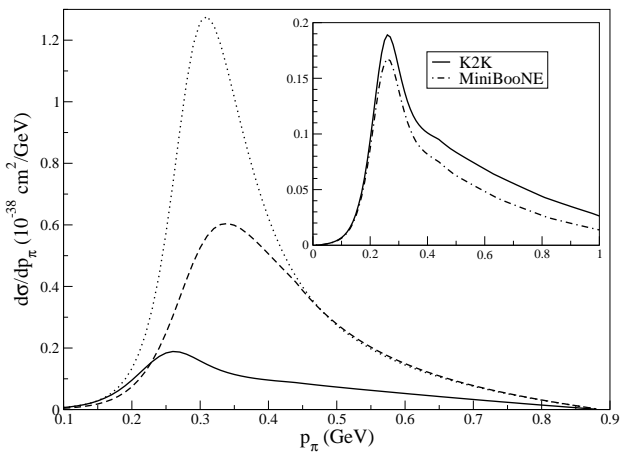}% Here is how to import EPS art
\caption{$\frac{d\sigma}{dp_\pi}~ vs~ p_\pi$ at $E_\nu$=1~GeV. The various curves correspond to the cases described in Fig.~4}
\end{figure}

The mass and width of the $\Delta$ appearing in eq.~(3) may be modified in the nuclear medium due to inhibition of the $\Delta$ decay caused by the Pauli blocking of nucleons in the $\Delta \rightarrow$N$\pi$ decay and due to two and three body $\Delta$ absorption processes like $\Delta$N$\rightarrow$NN, $\Delta$N$\rightarrow\Delta$N$\pi$ and $\Delta$NN$\rightarrow$NNN, which may take place in the nuclear medium. These modifications in the $\Delta$ properties have been studied by many authors~\cite{oset1},~\cite{oset2} by calculating the self energy $\Sigma_\Delta$ of $\Delta$ in the nuclear medium leading to density dependent changes in $\text M_\Delta$ and $\Gamma$, given by $\widetilde M_\Delta= \text M_\Delta + \text {Re}\Sigma_\Delta$ and $\frac{\widetilde\Gamma}{2}= \frac{\bar\Gamma}{2} - \text{Im}\Sigma_\Delta$, where $\bar\Gamma$ is the $\Delta$ width modified due to Pauli blocking. We use the expressions of Oset and Salcedo~\cite{oset1} for $\widetilde\Gamma$, ${\text Re}\Sigma_\Delta$ and ${\text Im}\Sigma_\Delta$  where the density dependence of $\Sigma_\Delta$ and $\widetilde\Gamma$ in the nuclear medium has been studied in a local density approximation to the delta hole model. Thus incorporating the nuclear medium effects, the $\Delta$-dependent hadronic factors in eq.~(3) become density dependent and the nuclear transition current ${J}^\mu_\text{W,Z}$ is now modified to
{\small
\begin{equation}
{J}^\mu_\text{W,Z}=\sum_\text{i=s,u}\int{\cal T}^{\mu}_\text{W}(i)\frac{\text M^2}{\text P^2_i-\widetilde{\text M}^2_{\Delta}+i\widetilde\Gamma \widetilde{\text M}_\Delta}\rho^i_\text{W,Z}(r)e^{i({\vec q}-{\vec p_\pi})\cdot{\vec r}}d{\vec r}
\end{equation}}

The final state interaction of the pion is taken into account by replacing the plane wave pion by a distorted wave pion. The distortion of the pion is calculated in the eikonal approximation in which the distorted pion wave function is written as
{\small
\begin{equation}
\widetilde\phi_\pi({\vec r})=e^{-i{\vec p_\pi}\cdot{\vec r}}~e^{-i\int^{\infty}_{z}\frac{1}{2p_\pi}\Pi(\rho({\vec b},z^\prime))dz^\prime}
\end{equation}}
where ${\vec r}=({\vec b},z)$. $\Pi(\rho)$ is the self energy of the pion calculated in the local density approximation of the delta hole model and is taken from Ref.~\cite{oset1}
{\small
\begin{equation}
\Pi(\rho)=\frac{4}{9}\left (\frac{f_{\pi \text N \Delta}}{\text m_{\pi}}\right)^{2}\frac{\text M^2}{\text W^2}|{\bf {\text p}_\pi}|^2\rho\frac{1}{\text W-\widetilde{\text M}_\Delta+\frac{i{\widetilde\Gamma}}{2}}
\end{equation}}

Note that we use the nonrelativistic form of the energy denominator in the delta propagator for calculating the pion self energy. 

The nuclear transition current operators ${J}^\mu_\text{W,Z}$ discussed above in eq.~(3), with incorporation of nuclear medium effects through eq.~(4) and the final state interaction effects through eq.~(5), are used to numerically evaluate the total cross sections $\sigma(\text E_\nu)$, differential cross sections $\frac{\text d\sigma}{\text dQ^2}$ and $\frac{\text d\sigma}{\text dp_\pi}$ for $^{12}\text{C}$, $^{27}\text{Al}$ and Freon(C$F_3$Br). We take the proton density $\rho_\text{p}(r)=\frac{Z}{A}\rho(r)$ and the neutron density $\rho_\text{n}(r)=\frac{A-Z}{A}\rho(r)$ which should be appropriate for describing nucleon densities in intermediate mass nuclei. The nuclear densities $\rho(r)$ are taken from Ref.~\cite{vries}. 

In Fig.~2, we have presented the results for the total scattering cross section ($\sigma^\text{CC}$) for the coherent charged current reaction induced by $\nu_\mu$ in $^{12}\text{C}$. The results for $\sigma^\text{CC}(\text E_\nu)$(scaled by a factor of $\frac{1}{2}$) vs $\text E_\nu$ are shown without nuclear medium effects and with nuclear medium effects. When the pion absorption and nuclear medium effects, are both taken into account the results for $\sigma^\text{CC}(\text E_\nu)$ are shown by the solid lines. We see that the nuclear medium effects lead to a reduction of 30-35$\%$ around $\text E_\nu$=1-1.5~GeV in $\sigma^\text{CC}(\text E_\nu)$ while the reduction due to final state interaction is quite large. We have also shown in this figure the results for $\sigma^\text{CC}(\text E_\nu)$ vs $\text E_\nu$ when a cut of 450~MeV is applied on the muon momentum as done in the K2K experiment. We obtain a total cross section of 6.93$\times 10^{-40}{\text{cm}^2}$ in $^{12}\text{C}$ which corresponds to 0.578$\times 10^{-40}\frac{\text{cm}^2}{nucleon}$, and is consistent with the experimental result of $\sigma^\text{CC}<$0.642$\times10^{-40}\frac{\text{cm}^2}{nucleon}$ in $^{12}\text{C}$ reported by the K2K collaboration~\cite{hasegawa}. Our calculations show that:

(i) the contribution to the cross section comes mainly from the s-channel diagram($>90\%$) which is dominated by on shell $\Delta$, thus making the off shell correction quite small. The inclusion of off shell effects by introducing a form factor at the $\pi N \Delta$ vertex~\cite{line}-\cite{post} leads to a reduction in the cross section which is estimated to be 4-6~\% in the energy region of 1-2~GeV.

(ii) the contribution to the cross section from the vector current is negligibly small ($<$2~\%) and the major contribution comes from the axial current only, leading to near equality of neutrino and antineutrino cross sections. 

 We show in Fig.~3, the total cross section $\sigma^\text{NC}(\text E_\nu)$ for the neutral current induced $\pi^0$ production from $^{12}\text{C}$, $^{27}\text{Al}$ and $\text{CF}_3\text{Br}$(Freon), along with the experimental results from the MiniBooNE collaboration for $^{12}\text{C}$~\cite{raaf}, from the Aachen collaboration for $^{27}\text{Al}$~\cite{faissner} and from the Gargamelle collaboration for Freon~\cite{isiksal}. We see that the theoretical results for the neutral current induced coherent $\pi^0$ production are in reasonable agreement with presently available experimental results in the intermediate energy region.

We also present in Figs. 4 and 5 the differential cross sections $\frac{\text d\sigma}{\text dq^2}$ and $\frac{\text d\sigma}{\text dp_\pi}$ for charged pion production at $\text E_\nu$=1~GeV where nuclear medium and final state interactions effects are shown explicitly. In the inset we exhibit the differential cross sections $<\frac{\text d\sigma}{\text dq^2}>$ and $<\frac{\text d\sigma}{\text dp_\pi}>$ averaged over the K2K and MiniBooNE neutrino spectra without applying any cuts on the muon momentum. 

We would  like to emphasize the important role of nuclear medium effects which reduce the total cross sections and help to obtain reasonable agreement with the experimental values. The uncertainty in calculating these nuclear effects comes mainly from  uncertainties in the value of the $\Delta$-self energy which is calculated~\cite{oset1} within an accuracy of 15-20~\%. This leads to an uncertainty of $8-12\%$ in the total cross section. There is an additional uncertainty due to various parameterizations used in the literature for the weak transition form factors~\cite{sakuda}, which is small (3-5~\%) as this process is dominated by differential cross sections at very low $q^2$(Fig.~4). Thus the total uncertainty in the present calculation is estimated to be about $15\%$.

Similar results for $\sigma^{\text CC,NC}({\text E}_\nu)$ in $^{12}\text{C}$ have been recently obtained by Paschos and Kartavtsev~\cite{paschos} who have calculated only the axial vector contributions to the total cross sections. They use Adler's PCAC theorem~\cite{adler} to relate the forward neutrino cross section to the $\pi$-$^{12}\text{C}$ elastic scattering and extend it to non-zero $Q^2$($\le 0.2$~GeV$^2$). The experimental data on $\pi$-$^{12}\text{C}$ elastic scattering cross sections have been used, which are available only in a limited region of pion energy and momentum transfer. Moreover, for other nuclei where no experimental cross sections are available an approximate scaling law for the A-dependence of pion-nucleus elastic scattering cross sections has been used to obtain cross sections for these nuclei from $^{12}$C data. In their model no nuclear effects are included in the weak pion production process and the final state interaction of pions is included only through the dynamics of pion-$^{12}$C elastic scattering. On the other hand, in the present work the results obtained from a microscopic study of the $\Delta$-nucleus and pion-nucleus dynamics have been used to evaluate various nuclear medium effects in the weak pion production process as well as in the final state interactions of pions with a nucleus.

To summarize, we have studied in this letter total cross section $\sigma(\text E_\nu)$, differential cross sections $\frac{\text d\sigma}{\text dq^2}$ and $\frac{\text d\sigma}{\text dp_\pi}$ for the neutrino induced coherent production of charged and neutral pions in a model, which takes into account nuclear medium effects in the weak pion production process through $\Delta$ dominance treated in local density approximation. The final state interaction of pions with the nucleus is described in an eikonal approximation with a pion optical potential derived in terms of the pion self energy in the nuclear medium. The good agreement between the theoretical and experimental results in the energy region of 1~GeV is obtained mainly due to inclusion of nuclear medium and pion absorption effects. The method may be useful to analyze the neutrino induced pion production data at neutrino energies relevant for neutrino oscillation experiments being done by K2K, MiniBooNE and J-PARC collaborations.

This work is financially supported by D.S.T., Govt. of India (Grant number DST-SP/S2/K-07/2000).


\begin{thebibliography}{2006}
\bibitem{line} While we use the on shell form of the $\Delta$ propagator, the off shell effects at $\Delta N\pi$ vertex has been studied using:
$f_ {\pi N \Delta}(P^2)=\frac{\Lambda^4}{\Lambda^4+(P^2-{M^2_\Delta})^2}f_ {\pi N \Delta}$ with $\Lambda=1$~GeV~\cite{post}.
\end{thebibliography}
\end{document}